\documentclass[conference]{IEEEtran}
\IEEEoverridecommandlockouts
\usepackage[top=0.7in, left=0.7in, right=0.7in, bottom=0.9in]{geometry}
\usepackage{amsmath,amsfonts}
\usepackage{algorithmic}
\usepackage{algorithm}
\usepackage{subcaption}

\usepackage{bbm}
\usepackage{graphicx}

\usepackage{xcolor}
\usepackage{float}   
\usepackage[normalem]{ulem}
\usepackage{dsfont}
\usepackage{textcomp}
\usepackage{stfloats}
\usepackage{url}
\usepackage{verbatim}
\usepackage{algorithmic}
\usepackage{array}
\usepackage{booktabs}
\usepackage{pifont}
\usepackage{comment}
\usepackage{algorithm}
\usepackage{algorithmic}
\usepackage[acronym]{glossaries}
\newacronym{ack}{ACK}{Acknowledgement}
\newacronym{awgn}{AWGN}{Additive white Gaussian Noise}
\newacronym{ai}{AI}{Artificial Intelligence}
\newacronym{aoi}{AoI}{Age of Information}
\newacronym{bs}{BS}{Base Station}

\newacronym{cts}{CTS}{Clear to Send}
\newacronym{cdf}{CDF}{Cumulative Distributed Function}
\newacronym{ccdf}{CCDF}{Complementary Cumulative Distributed Function}
\newacronym{dl}{DL}{Deep Learning}
\newacronym{dnn}{DNN}{Deep Neural Networks}
\newacronym{dram}{DRAM}{Dynamic Random Access Memory}

\newacronym{eo}{EO}{Earth Observation}
\newacronym{es}{ES}{Edge Server}

\newacronym{fov}{FoV}{Field of View}
\newacronym{fp}{FP}{False Positive}
\newacronym{fn}{FN}{False Negative}

\newacronym{gs}{GS}{Ground Station}

\newacronym{id}{ID}{Identity}
\newacronym{idwu}{IDWu}{Identity-based Wake-up}
\newacronym{iot}{IoT}{Internet of Things}

\newacronym{kpis}{KPIs}{Key Performance Indicator}
\newacronym{kkt}{KKT}{Karush-Kuhn-Tucker Conditions}

\newacronym{mcu}{MCUs}{Micro Controller Unit}
\newacronym{mac}{MAC}{Medium Access Control}
\newacronym{muac}{MUAC}{Multiply-accumulation}
\newacronym{mcwu}{MCWu}{MultiCast Wake-up}
\newacronym{madd}{mADD}{multiplication and addition}
\newacronym{ml}{\textrm{ML}}{Machine Learning}
\newacronym{mlmo}{ML-Mode}{\gls{ml} Sharing Mode}
\newacronym{mlmosg}{MLMoSg}{\gls{mlmo} Signal}
\newacronym{ook}{OOK}{On-Off Keying}
\newacronym{pmf}{PMF}{Probability Mass Function}
\newacronym{pdf}{PDF}{Probability Density Function}
\newacronym{qnns}{QNNs}{Quantized Neural Networks}

\newacronym{ppp}{PPP}{Poisson point process}
\newacronym{tinyairnet}{\textrm{TinyAirNet}}{Tiny Neural Network transmission over the Air}
\newacronym{ri}{RI}{Receiver-Initiated}
\newacronym{rimosg}{RIMoSg}{RI-Mode Signal}
\newacronym{semdas}{SEMDAS}{SEMantic DAta Soucing}
\newacronym{sram}{SRAM}{Static Random Access Memory}
\newacronym{sifi}{SiFi}{Significance and Fidelity}
\newacronym{ti}{TI}{Transmitter-Initiated}
\newacronym{timosg}{TIMoSg}{TI-Mode Signal}

\newacronym{tp}{TP}{True Positive}
\newacronym{tn}{TN}{True Negative}
\newacronym{jscc}{JSCC}{Joint Source Channel Coding}

\newacronym{uav}{UAV}{Unmannded  Aerial Vehicle}
\newacronym{rr}{RR}{Round-robin}

\newacronym{wsn}{WSNs}{Wireless Sensor Networks}
\newacronym{wmsn}{WMSNs}{Wireless Memory Sensor Networks}

\newacronym{wurx}{WuRx}{Wake-up receiver}
\usepackage{amsmath,amsfonts, amssymb}
\usepackage{bm}
\usepackage{graphicx}
\usepackage{cite}

\usepackage[font=small]{caption}
\usepackage[font=footnotesize]{subcaption}
\usepackage{tikz}
\usetikzlibrary{plotmarks,patterns,decorations.pathreplacing,backgrounds,calc,arrows,arrows.meta,spy,matrix}
\usepackage{tikzscale}

\usepackage{pgfplots}
\usepgfplotslibrary{colorbrewer, groupplots, patchplots}
\pgfplotsset{
    compat=newest,
    legend style={font=\footnotesize, fill opacity=0.7,  draw opacity=1, text opacity=1, draw=white!15!black, legend cell align=left, align=left}, 
    width=0.8\columnwidth, 
    scale only axis,
    height=4cm,
    yminorticks=false,
    xminorticks=false,
    label style={font=\small},
    title style={font=\small},
    tick align=outside,
    tick pos=left,
    tick style={color=black},
    tick label style={font=\footnotesize},
    grid style={line width=.1pt, draw=gray!20},
    major grid style={line width=.1pt,draw=gray!20},
    plot coordinates/math parser=false 
}
\newlength\figureheight
\newlength\figurewidth

\usepackage{multirow}
\usepackage{tablefootnote}
\usepackage{booktabs}
\usepackage{tabularx}

\def\BibTeX{{\rm B\kern-.05em{\sc i\kern-.025em b}\kern-.08em T\kern-.1667em\lower.7ex\hbox{E}\kern-.125emX}}

\newcommand{\probP}{\text{I\kern-0.15em P}}

\definecolor{amaranth}{rgb}{0.9, 0.17, 0.31}

\begin{document}

\title{Wireless Memory Approximation for Energy-efficient Task-specific IoT Data Retrieval}
\author{Junya Shiraishi,~Shashi Raj Pandey, Israel~Leyva-Mayorga, and~Petar Popovski\\
Department of Electronic Systems, Aalborg University. \\Email: \{jush, srp, ilm, petarp\}@es.aau.dk.
\thanks{
The work of J. Shiraishi was supported by European Union's Horizon Europe research and innovation funding programme under Marie Sk{\l}odowska-Curie Action (MSCA) Postdoctoral Fellowship, ``NEUTRINAI" with grant agreement No.~101151067. The work of P. Popovski was supported by the Velux Foundation, Denmark, through the Villum Investigator Grant WATER, nr. 37793. }
}
\maketitle
\begin{abstract}
    
      The use of Dynamic Random Access Memory (DRAM) for storing Machine Learning (ML) models plays a critical role in accelerating ML inference tasks in the next generation of communication systems. However, periodic refreshment of DRAM results in wasteful energy consumption during standby periods, which is significant for resource-constrained Internet of Things (IoT) devices. To solve this problem, this work advocates two novel approaches: 1) wireless memory activation and 2) wireless memory approximation. These enable the wireless devices to efficiently manage the available memory by considering the timing aspects and relevance of ML model usage; hence, reducing the overall energy consumption. Numerical results show that our proposed scheme can realize smaller energy consumption than the always-on approach while satisfying the retrieval accuracy constraint. 
\end{abstract}

 \section{Introduction}
\IEEEPARstart{T}{he} use of lightweight, on-device \gls{ai}/\gls{ml} models can empower \gls{iot} devices to transmit only data that is critical for the specific downstream tasks~\cite{chen2025toward}. This allows each \gls{iot} device to reduce its energy consumption. For example, introducing Tiny\gls{ml} can significantly reduce energy waste of \gls{iot} devices by letting them filter out irrelevant data transmissions~\cite{shiraishi2023energy}. Related work  on goal-oriented communications, such as task-specific \gls{iot} data retrieval and semantic queries~\cite{shiraishi2023energy}, largely focuses on designing energy-efficient communication protocols by running \gls{ai}/\gls{ml} algorithms on devices for the semantic extraction and interpretation of data locally. 
While memory management is not commonly considered a factor for energy consumption, it can have a major impact on \gls{iot} devices. This is especially true in the context of intelligent 6G networks, where the \gls{iot} devices have intelligent modules, such as Tiny\gls{ml}, and their transmissions are conditioned on the inferences carried out by these modules~\cite{shiraishi2023energy}.

\Gls{dram} technology is used in a wide range of systems, including, intelligent \gls{iot} camera systems for real-time operation. 
During the operation of such systems, the energy consumption of \gls{dram} can be divided into two parts:  1) data storage and access, and 2) memory refreshment. Specifically, throughout a standby period, \gls{dram} needs a periodic refreshment due to its volatility. This refreshment dominates the energy consumption of the device~\cite{liu2011flikker,bhati2015dram}, which consumes up to 50$\%$ of total energy consumption of \gls{dram}~\cite{nguyen2020approximate}.  
Importantly, the device should keep refreshing its memory even if in the absence of a computation requirement. This is necessary to be able to swiftly respond to the arrival of a computational task quickly. The energy waste due to standby-mode memory refreshments has received relatively low attention in the research literature, as compared to the energy reduction technology during the idle-period of communication~\cite {piyare2017ultra}. 

Memory approximation is an attractive approach to reduce energy consumption~\cite{leon2025approximate,nguyen2020approximate}, in which the \gls{dram} refresh rate is reduces, lowering the energy consumed by the refreshment at the cost of deterioration of accuracy. 
In~\cite{liu2011flikker}, the memory controller adapts the refreshing time considering data importance or criticality.     
In~\cite{wang2018content}, content-aware refresh has been proposed aiming to reduce the total number of times that the memory is refreshed. 
The concept of approximation of \gls{dram} memory is investigated for smart camera systems, considering the cost of communication, sensing, processing, and memory access~\cite{ghosh2023energy}. However, it does not consider how to reduce the wasteful memory access during the stand-by period, considering the timing and relevance to the current tasks of the receivers. Using memory irrespective of its importance for the current task is clearly a waste of energy. 
In order to solve the above-mentioned problem, this paper brings two concepts inspired by goal-oriented communication~\cite{strinati2024goal}: \emph{i}) \emph{wireless memory activation} and \emph{ii}) \emph{wireless memory approximation} (c.f. Sec.~\ref{sec:Wireless_Memory}). The fundamental principles of both concepts are to use memory only at right timing and only for the relevant model, as summarized follow:
\begin{itemize}
\item{Wireless memory activation: Realizing remote activation of memory of \gls{iot} device at the specific timing.}
\item{Wireless memory approximation: Remotely approximating memory use by specifying the refresh period of \gls{dram} memory considering the importance level of specific \gls{ml} model to serve the requested tasks.}
\end{itemize}
These concepts enable the \gls{iot} device to spend energy at the memory only when the specific model is highly likely required to complete the future task, leading to the significant energy reduction for memory use during the stand-by period.

To formalize and validate the concepts of wireless memory activation and approximation, this work focuses on a particular low-latency remote inference scenario, where the \gls{es} is interested in retrieving task-specific data with low latency to serve the user's requirement. The specific applications include wireless image retrieval tasks in \gls{iot} networks as in~\cite{shiraishi2023energy}.  
Our contributions can be summarized as follows: \emph{i}) We propose wireless memory activation and approximation for energy-efficient task-specific \gls{iot} data retrieval; \emph{ii}) We introduce a pull-based communication framework and design an associated energy-efficient communication/memory access protocol that takes into account the timing of memory use for the relevant \gls{ml} model; \emph{iii}) We characterize the system-level performance of the proposed scheme and elicit the gain in terms of total energy consumption and retrieval accuracy.

\section{System Model}\label{sec:system} 
We consider a simplified setup to elucidate the memory activation/approximation problem. The network consists of an \gls{es}, a single \gls{iot} device equipped with vision sensor, a radio interface, memory controller, \gls{dram}, and a storage, and an external service requester (user). Each of these entities communicates over a wireless medium.  
We consider a pull-based communication system in which the \gls{es} is conducting a specific task of data collection from the \gls{iot} device, as requested by the external user.

The external user requests the data acquisition task to the \gls{es}, which follows a Poisson process with the average arrival rate $\lambda$ [query/s]. 
We define the time of the previous query arrival as 0 and use $t_q$ to denote the random variable that represents the current query instance (episode). To facilitate our analysis of the gain brought by our method (c.f. Sec.~\ref{sec:Wireless_Memory}), this work focuses on the behavior of memory on the specific instance of the task-specific \gls{iot} data retrieval, i.e., the memory behavior of \gls{iot} device during the duration of [$0, t_q $].  

Upon receiving this query request from the external user, the \gls{es} forwards a query to the target \gls{iot} device. The \gls{iot} device detecting this query, processes and transmits the data, such as the estimated label obtained by \gls{ml} model, to the \gls{es}. We further assume an error-free channel for both uplink and downlink communication.

Furthermore, we assume that the \gls{es} has perfect knowledge about $\lambda$ based on the data collected in the past, but the actual query timing is unknown. 
The goal of this work is to design an energy-efficient data transmission framework that facilitates timely acquisition of only the relevant data, defined through user query, while considering the overall cost of memory usage. This involves wireless remote memory activation and approximation procedure, as illustrated in the system overview in Fig.~\ref{Fig:wireless_memory}, and discussed next. 
\subsection{Wireless Memory Activation and Approximation}\label{sec:Wireless_Memory}
\begin{figure}[t]
\centering
\includegraphics[width=0.48\textwidth]{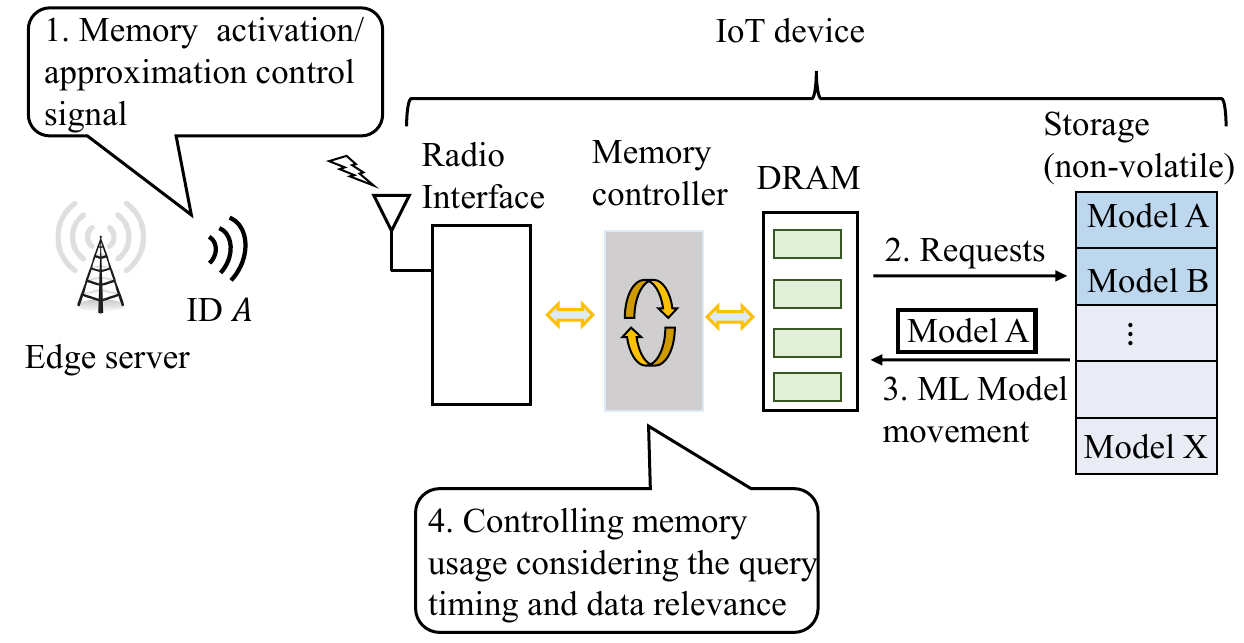}
\caption{The overview of the system integrating remote memory activation/approximation.}
\vspace{-4mm}
\label{Fig:wireless_memory}
\end{figure}

Now, we introduce the concept of wireless memory activation and approximation. The basic idea is illustrated in Fig.~\ref{Fig:wireless_memory}.  
When a new episode begins at $t =0$, the \gls{es} first decides the timing of the transmission of the memory activation signal based on the statistical knowledge that the \gls{es} has about the query arrival for the specific task, i.e., average arrival rate $\lambda$. Specifically, the \gls{es} transmits a memory activation signal at $\zeta$ [s] after the current episode begins. The memory activation signal includes the information of the specific model to be moved to the memory related to the current task, e.g, by embedding the model \gls{id} information into the control signal (e.g. model \gls{id} $A$). 
If the receiver of \gls{iot} device detects this control signal, it checks whether it satisfies the model activation conditions.  
Specifically, it checks whether information of model \gls{id} embedded into control signal coincides with the model \gls{id} stored in the storage. If so, it moves the specific model $\mathcal{M}_A$ from the storage to the memory.
Then, each device initiates processing using the \gls{ml} model. This enables demand-driven \gls{dram} activation, reducing wasteful energy consumption during the stand-by periods. After sending the memory activation signal, the \gls{es} sends the memory approximation signal so that it can set the refresh period to be $\tau$. Longer $\tau$ means that \gls{es} can accept higher information loss, that is, coarser data approximation. 
The wireless memory approximation aims to remotely control the refresh period of memory based on the importance of each model in terms of expected future demand from the external user. For example, when the \gls{es} knows the data point $X$ (such as feature map) is more important than the data point $Y$, it is reasonable to allocate a small (large) refresh period for data point $X$ ($Y$) to efficiently control the efficient memory use. To this end, one could send a control signal or a behavior model so that the \gls{iot} device can autonomously tune the memory approximation based on the installed model. This work focuses on the former approach. 

The total overhead required for the wireless memory activation and approximation is $T_{\mathrm{\Sigma}} = T_{W} +T_{\mathrm{approx}}+T_{l}$, where $T_{W}$, $T_{\mathrm{approx}}$, and $T_{l}$, respectively, represent the time required for control signal transmission, the additional overhead required for notifying the approximation of the model, and the time required for loading the model.

\subsection{\gls{dram} Retention Accuracy}\label{eq:dram_model}  
\gls{dram} must be continuously refreshed to retain the data, as the content is volatile due to capacitor discharge.  The refresh interval $\tau$ is a key to control the quality of data in the memory, i.e., to control the level of approximation of \gls{dram}.  
A larger refresh interval $\tau$ can reduce the energy use of memory, while it increases the chances of bit errors in \gls{dram}~\cite{raha2016quality}, as capacitors may discharge below a threshold needed to retain information~\cite{ghosh2023energy}.  
Normally, \gls{dram} is responsible to store the images and different \gls{dnn} parameters, including weights and feature maps~\cite{ghosh2023energy}. The focus of this work is to characterize the accuracy of \gls{ml} model through the different refresh rate and the timing of memory activation. 
To enable the adaptation of different refresh interval, it needs to introduce different quality bins $\mathcal{Q}= \{q_1, q_2, \ldots, q_{I_{\mathrm{M}}}\}$, each of which is allocated a different refresh period. Here, a larger quality bins represents a higher refresh interval, leading to a higher bit error. Then, each data point, e.g., \gls{ml} model for the specific task will be assigned to one of the quality bin in $\mathcal{Q}$, based on the mapping function.

To characterize the accuracy degradation through the parameter of $\tau$ and $\zeta$, we model the state of each memory array with two state Markov chain, consisting of correct state and erroneous state. Here, each memory array is assumed to be in the correct state when it is loaded. 
In each refreshment, the state transition from the correct state to error state happens with probability $P_e(\tau)$, depending on the refresh period $\tau$:   
\begin{equation}
 P_e(\tau) = 1 -\exp\left(-\frac{\tau-\tau_{\mathrm{0}}}{\beta}\right),\label{eq:bit_error_against_tau}
\end{equation}
where $\tau_{\mathrm{0}} = 64$ [ms] is the refresh time required to maintain the data in \gls{dram} without error and $\beta$ is a hardware  specific constant value representing the degree of accuracy degradation; smaller (larger) $\beta$ corresponds to the case where the speed of memory degradation is higher (lower).  
On the other hand, we model that the error state is an absorbing state, meaning that after entering an erroneous state, it no longer returns to the correct state. 
Given the refresh period $\tau$ and total memory activation period $Q = t_q - \zeta$, the total number of refreshments during a single episode is $M_{\zeta} = \lfloor \frac{Q}{\tau} \rfloor$. 
Then, the probability that the state is in the erroneous state after $M_{\zeta}$ refreshments can be expressed as: 
\begin{equation}
 \bar{P}_{e}(M_{\zeta}, \tau) = 1 -\left(1 -P_e(\tau)\right)^{M_{\zeta}}. \label{eq:bit_error_after_refreshment}
\end{equation}

In this work, to facilitate our analysis, we assign a unique refresh period $\tau$ for all data points $\bm{x}= [x_1, x_2, \ldots, x_{N}]^{\top}$ to analyze the system-level trade-off between accuracy and energy, where $N$ is the total number of bits stored in the \gls{dram}. 
In practice, the refresh period $\tau$ should be decided given data point ${\bm{x}}$ considering the maximum allowable distortion $\Delta~\in [0, 1]$. To define allowable distortion $\Delta$, this paper considers Hamming distortion. Let ${{\bm{y}}} = [y_1, y_2, \ldots, y_{N}]^\top$ be the decayed data point of ${\bm{x}}$. Then, the Hamming distortion between ${\bm{x}}$ and ${\bm{y}}$ is expressed as: 
 \begin{equation}
  d({\bm{x}}, {\bm{y}}) = \frac{1}{N} \sum_{j=1}^{N} \mathbbm{1} (x_j \neq y_j).
 \end{equation}
 Finally, we define \emph{retention accuracy}, the accuracy threshold that the \gls{ml} model can retain in memory, as $d({\bm{x}}, {\bm{y}}) \leq \Delta$, based on the predefined threshold $\Delta$. 
Accordingly, we can derive \emph{retention accuracy}, which is the probability that the model can provide correct output after $M_{\zeta}$ refreshments based on Hamming distortion. Formally, given $\Delta$ and $Q$, the retention accuracy, defined as $\gamma(\tau, Q)  = \Pr\left(d\left({\bm{x}}, {\bm{y}}\right) \leq \Delta\right)$ can be expressed as: 
\begin{equation}
 \gamma(\tau, Q)=\sum_{k=0}^{\lfloor{N\Delta}\rfloor}\binom{N}{k}\bar{P}_{e}\left(M_{\zeta}, \tau\right)^k\left(1-\bar{P}_{e}(M_{\zeta}, \tau)\right)^{N-k}.\label{eq:retention_accuracy}  
\end{equation}

\subsection{Retrieval Accuracy}
The retrieval accuracy, i.e., the accuracy of received data with respect to the specific task can be defined by considering two aspects: 1) whether the \gls{es} successfully collects data from the \gls{iot} device and 2) the accuracy of output of \gls{ml} model in Eq.~\eqref{eq:retention_accuracy}.  
Formally, we define the retrieval accuracy as follows: 
\begin{equation}
  U( \tau, \zeta, T_{\Sigma})= h(\zeta)\gamma(\tau, Q),\label{eq:Utility}
\end{equation}
where $h(\zeta) \in \{0, 1\}$ is a Bernoulli random variable, representing whether the \gls{es} successfully received task-specific data from the \gls{iot} device upon the query request. Under the error-free channel, the \gls{es} can successfully receive the data related to the current task only if the memory is activated by the time $t_q$. Here, $h(\zeta) = 1$ only if it satisfies $(\zeta + T_{\Sigma}) < t_{q}$; otherwise $h(\zeta) = 0$. 
On the other hand, the accuracy of \gls{ml} model $\gamma(\tau, Q)$ depends on the local inference accuracy, influenced by the refresh period $\tau$ and $Q$.

\subsection{Energy Model}\label{sec:energy_model}
The total energy consumption of an \gls{iot} device, $E_{\mathrm{tot}}(t_q)$ can be calculated by taking into account the energy consumed for \emph{i}) communication, $E_{\mathrm{c}}(t_q)$, \emph{ii}) computation, $E_{\mathrm{p}}(t_q)$, \emph{iii}) refreshment, $E_{\mathrm{re}}(t_q)$, and \emph{iv}) \gls{ml} model loading from the storage to memory, $E_{\mathrm{l}}(t_q)$, during a single episode. 
 
The energy consumed for refreshment depends on the total number of refreshment during a single episode, denoted as $n_{\mathrm{re}}(\tau, \lambda)$, which is a function of $\lambda$ and the refresh period $\tau$. Then, let $b_{\mathrm{row}}, I_M$, and $\xi_{\mathrm{re}}$ be the data size that can be stored in a single row in \gls{dram}, the total number of bins that the \gls{iot} device uses during a single episode, and the energy consumption for the refreshment of \gls{dram} row, respectively. Then, the total energy consumed for refreshment becomes:
\begin{equation} 
E_{\mathrm{re}}(t_q) =  \xi_{\mathrm{re}}n_{\mathrm{re}}(\tau, \lambda)I_Mb_{\mathrm{row}}.
\end{equation}
On the other hand, we model the energy required by the other units as: $E_{\mathrm{c}}(t_q) + E_{\mathrm{p}}(t_q) = w_{c}I_{\mathrm{M}}\xi_{\mathrm{re}}$ and $E_{\mathrm{l}}(t_q) = w_{l}I_{\mathrm{M}}\xi_{\mathrm{re}}$. Here, $w_{c}$ and $w_{l}$ are, respectively, a scaring factor controlling the ratio of energy consumption of $E_{\mathrm{c}}(t_q) + E_{\mathrm{p}}(t_q)$ and  $E_{\mathrm{l}}(t_q)$ against the energy consumed for the single refreshment. 

\section{Analysis and Problem Formulation}\label{sec:analysis_problem}
\begin{figure}[t]
\centering
\includegraphics[width=0.48\textwidth]{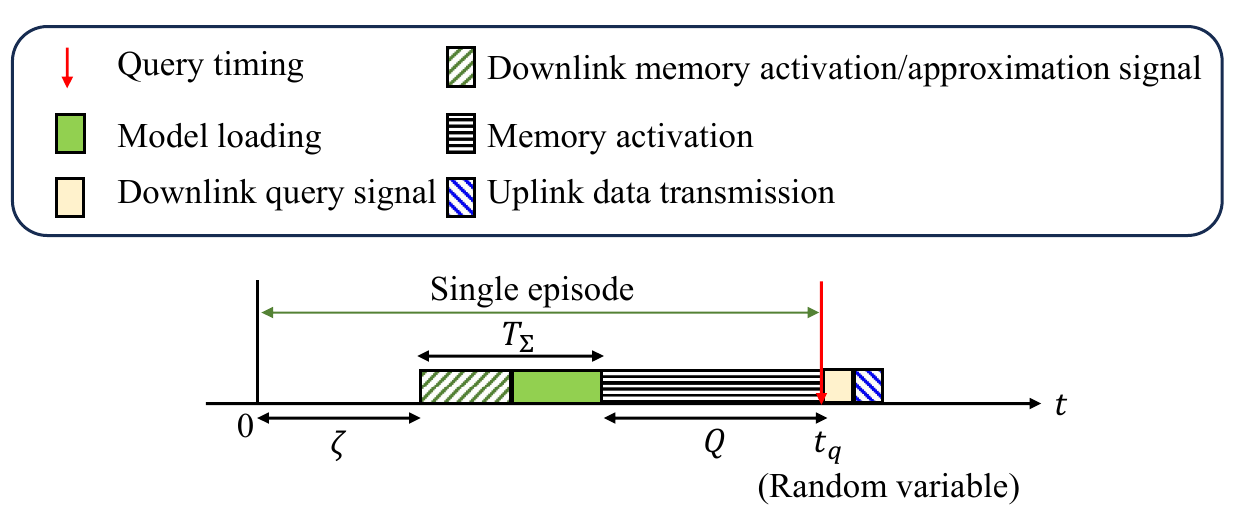}
\caption{Analytical model.}
\vspace{-4mm}
\label{Fig:Analytical_model}
\end{figure}

The retrieval accuracy and total energy consumption of the proposed scheme is derived based on the the analytical model illustrated in Fig.~\ref{Fig:Analytical_model}.  

Considering the \gls{pdf} of query arrival time and the definition of retrieval accuracy defined in~Eq.~\eqref{eq:retention_accuracy}, the expected retrieval accuracy of our proposed scheme can be expressed as follows:
\begin{equation}\begin{split}
&\mathbb{E}[U^{\mathrm{p}}( \tau, \zeta, T_{\Sigma})]=\sum_{k=0}^{\lfloor{N\Delta}\rfloor} \Pr(N_e(M_{\zeta}) = k) \\& =e^{-\lambda\,\left(\zeta +T_{\Sigma}\right)}\sum_{k=0}^{\lfloor{N\Delta}\rfloor} \left(1-\psi\right)\binom{N}{k}\sum_{j=0}^{k}\frac{\binom{k}{j}(-1)^j}{1-\psi\,q_e^{c}},\label{eq:PMF_bit_error}
\end{split}\end{equation}
where $\psi = \exp(-\lambda\,\tau)$, $c = N-k+j$, and $q_e = 1-P_e$. 

On the other hand, based on the energy model mentioned in Sec.~\ref{sec:energy_model}, the total energy consumption of the proposed scheme during a single episode can be expressed as follows: 
\begin{equation}
   \mathbb{E}[E_\mathrm{tot}^\mathrm{p}]=e^{-\lambda\,\left(\zeta +T_{\Sigma}\right)}\left(\frac{\psi}{1-\psi} +w_{c}+w_{l}\right) I_{\mathrm{M}}\xi_{\mathrm{re}}.\label{eq:proposed_ene}
\end{equation}
See the Appendix for the detailed derivation of Eqs.~\eqref{eq:PMF_bit_error}--~\eqref{eq:proposed_ene}.

Considering the trade-off between the retrieval accuracy and total energy consumption through the values of $\{\zeta, \tau\}$, here, we evaluate the minimum energy consumption under the retrieval accuracy constraint $U_{\mathrm{th}}$, formulated as: 
\begin{equation}
\underset{\{\tau, \zeta\}}{\min}~\mathbb{E}[E_\mathrm{tot}^\mathrm{p}]~\text{s.t.}~\mathbb{E}[U^{\mathrm{p}}]~\geq~U_{\mathrm{th}}.\label{eq:object function}
\end{equation}
Here, as the problem in Eq.~\eqref{eq:object function} is non-convex, we rely on the grid-search method to find the optimal solution. Specifically, we select the value of $\zeta$ in the range from 0 to 250 with a step of 0.5, and that of $\tau$ in the range from 64 ms to 2560 ms with a step of 64 ms.

\section{Numerical Evaluations}
In this section, we present results for the retrieval accuracy and the normalized energy consumption for the different parameters. These results were obtained through Monte Carlo simulation, with over $10^4$ trials, based on the description of Sec.~\ref{sec:system} and obtain the results of the analysis described in Sec.~\ref{sec:analysis_problem}.  

As a baseline scheme, we consider the always-on approach, in which \gls{dram} remains active holding the task-specific \gls{ml} model during the entire episode, in which the memory is refreshed with periodicity $\tau_0$. For the sake of brevity, we omit the detail here, but the retrieval accuracy and total energy consumption of baseline scheme can be expressed as $\mathbb{E}[U^{\mathrm{b}}] = 1$ and $\mathbb{E}[E_\mathrm{tot}^\mathrm{b}] = \left( \frac{ \psi_0}{1-\psi_0} +w_{c}\right)I_{\mathrm{M}} \xi_{\mathrm{re}}$, where $\psi_0 = e^{-\lambda\,\tau_0}$. 
Here, the normalized energy consumption of the proposed scheme can be defined as $\bar{E}={\mathbb{E}[E_\mathrm{tot}^\mathrm{p}]}/{\mathbb{E}[E_\mathrm{tot}^\mathrm{b}]}$, which can be expressed as follows:
\begin{equation}
\bar{E}=
e^{-\lambda\,\left(\zeta +T_{\Sigma}\right)}\frac{(1-\psi_0)\psi + (w_{c}+w_{l})(1-\psi_0)(1-\psi)}{\psi_0(1-\psi) + (1-\psi_0)(1-\psi)w_{c}}.\label{eq:Normalized_ene} 
\end{equation}

\subsection{Numerical Results}
\subsubsection{Behavior of \gls{dram} Retention Accuracy Model}
\begin{figure}[t!]
\begin{subfigure}[h]{0.2\textwidth}
\centering
\begin{tikzpicture}[scale=1]

\begin{axis}[%
height=3.5cm,
scale only axis,
xmin=50,
xmax=250,
xtick={50,100,150,200,250},
 xticklabels={
 \(\displaystyle 50\),
  \(\displaystyle 100\),
  \(\displaystyle 150\),
   \(\displaystyle 200\),
  \(\displaystyle 250\),
  },
xlabel style={font=\color{white!15!black}},
xlabel={$Q$~[s]},
ymin=0,
ymax=1.1,
ylabel style={font=\color{white!15!black}},
ylabel={$\gamma(\tau, Q)$},
axis background/.style={fill=white},
legend style={at={(0.22,0.9)},anchor=north west,legend cell align=left, align=left, draw=none,fill=none },
 legend image post style={xscale=0.8},
legend columns=1
]
\addplot[solid, color=red, line width=1pt, mark size=1pt]
  table[row sep=crcr]{%
25	1\\
50	1\\
75	1\\
100	1\\
125	1\\
150	1\\
175	1\\
200	1\\
225	1\\
250	1\\
};
\addlegendentry{$\tau = 64$ ms}
\addplot [color=red, only marks, mark=square, mark options={solid, red}, forget plot]
  table[row sep=crcr]{%
25	1\\
50	1\\
75	1\\
100	1\\
125	1\\
150	1\\
175	1\\
200	1\\
225	1\\
250	1\\
};
\addplot[dashed, color=blue, line width=1pt, mark size=1pt]
  table[row sep=crcr]{%
25	0.999762609272245\\
50	0.988942317700791\\
75	0.929525610738677\\
100	0.796366491479784\\
125	0.614651891004524\\
150	0.429674476897165\\
175	0.27480057247704\\
200	0.163655874602711\\
225	0.0916144103948964\\
250	0.0485057478519858\\
};
\addlegendentry{$\tau = 128$ ms}
\addplot [color=blue, only marks, mark=diamond, mark options={solid, blue}, forget plot]
  table[row sep=crcr]{%
25	0.999\\
50	0.99\\
75	0.921\\
100	0.799\\
125	0.604\\
150	0.416\\
175	0.282\\
200	0.157\\
225	0.082\\
250	0.05\\
};
\addplot[dotted, color=brown, line width=1pt, mark size=1pt]
  table[row sep=crcr]{%
25	0.998733464494701\\
50	0.957203299090471\\
75	0.797208936857203\\
100	0.552197733299719\\
125	0.321540069713027\\
150	0.163655874602762\\
175	0.0746001361639704\\
200	0.0311297461816668\\
225	0.012091636089855\\
250	0.00439276862615606\\
};
\addlegendentry{$\tau = 192$ ms}
\addplot [color=brown, only marks, mark=pentagon, mark options={solid, brown}, forget plot]
  table[row sep=crcr]{%
25	0.999\\
50	0.952\\
75	0.806\\
100	0.529\\
125	0.321\\
150	0.15\\
175	0.085\\
200	0.023\\
225	0.014\\
250	0.004\\
};

\end{axis}
\end{tikzpicture}%
\label{subFig:gamma_against_Q}
\end{subfigure}
\hspace{1em}
\begin{subfigure}[h]{0.2\textwidth}
\centering
\begin{tikzpicture}[scale=1]

\begin{axis}[%
height=3.5cm,
scale only axis,
xmin=0.025,
xmax=0.25,
xtick={0.05,0.1,0.15,0.2,0.25},
 xticklabels={
 \(\displaystyle 0.05\),
  \(\displaystyle 0.1\),
  \(\displaystyle 0.15\),
   \(\displaystyle 0.2\),
  \(\displaystyle 0.25\),
  },
xlabel style={font=\color{white!15!black}},
xlabel={$\Delta$},
ymin=0,
ymax=1,
ylabel style={font=\color{white!15!black}},
ylabel={$\gamma(\tau, Q)$},
axis background/.style={fill=white}
axis background/.style={fill=white},
legend style={at={(0.3,0.42)},anchor=north west,legend cell align=left, align=left, draw=none,fill=none },
 legend image post style={xscale=0.8},
legend columns=1
]
\addplot[solid, color=red, line width=1pt, mark size=1pt]
  table[row sep=crcr]{%
0.025	0.0420554596539958\\
0.05	0.283240153671155\\
0.075	0.47013677689849\\
0.1	0.797208936857203\\
0.125	0.951147224138236\\
0.15	0.979608520639117\\
0.175	0.997364343461674\\
0.2	0.999176265861783\\
0.225	0.999937968216936\\
0.25	0.999996624430761\\
};
\addlegendentry{$\beta = 320$}
\addplot [color=red, only marks, mark=square, mark options={solid, red}, forget plot]
  table[row sep=crcr]{%
0.025	0.04\\
0.05	0.2781\\
0.075	0.465\\
0.1	0.791\\
0.125	0.9516\\
0.15	0.977\\
0.175	0.9969\\
0.2	0.999\\
0.225	0.9999\\
0.25	1\\
};
\addplot[dashed, color=blue, line width=1pt, mark size=1pt]
  table[row sep=crcr]{%
0.025	0.158380497342506\\
0.05	0.591598007164909\\
0.075	0.776556589827224\\
0.1	0.957203299090467\\
0.125	0.995160345217079\\
0.15	0.998633598730665\\
0.175	0.999920400155669\\
0.2	0.999983383636728\\
0.225	0.999999445465596\\
0.25	0.999999986708208\\
};
\addlegendentry{$\beta = 480$}
\addplot [color=blue, only marks, mark=diamond, mark options={solid, blue}, forget plot]
  table[row sep=crcr]{%
0.025	0.1603\\
0.05	0.5891\\
0.075	0.7798\\
0.1	0.9557\\
0.125	0.9958\\
0.15	0.9989\\
0.175	0.9996\\
0.2	1\\
0.225	1\\
0.25	1\\
};
\addplot[dotted, color=brown, line width=1pt, mark size=1pt]
  table[row sep=crcr]{%
0.025	0.292183243369409\\
0.05	0.770971337831559\\
0.075	0.901972536458179\\
0.1	0.988942317700791\\
0.125	0.999282303345473\\
0.15	0.999847268911494\\
0.175	0.999994975388749\\
0.2	0.999999213408575\\
0.225	0.999999985276742\\
0.25	0.999999999802566\\
};
\addlegendentry{$\beta = 640$}
\addplot [color=brown, only marks, mark=pentagon, mark options={solid, brown}, forget plot]
  table[row sep=crcr]{%
0.025	0.2904\\
0.05	0.7661\\
0.075	0.9052\\
0.1	0.9883\\
0.125	0.9992\\
0.15	0.9999\\
0.175	1\\
0.2	1\\
0.225	1\\
0.25	1\\
};
\end{axis}
\end{tikzpicture}%
%
\label{subFig:gamma_against_RT_th}
\end{subfigure}

\caption{Retention accuracy defined in Eq.~\eqref{eq:retention_accuracy} as a function of memory activation period $Q$ (left figure) and $\Delta$ (right figure) for the results obtained by Eq.~\eqref{eq:retention_accuracy} (line) and by simulations (marker). }
 \vspace{-4mm}
 \label{Fig:retention_analysis}
\end{figure}
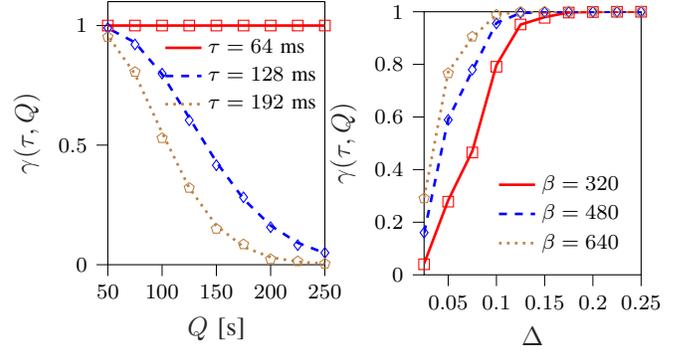

Fig.~\ref{Fig:retention_analysis} (left) shows retention accuracy in Eq.~\eqref{eq:retention_accuracy} as a function of memory activation period $Q$, where we set $N = 64$, $\beta = 640$ and $\Delta = 0.1$. From this figure, we can see that the retention accuracy $\gamma(\tau, Q)$ becomes small as refresh period $\tau$ becomes large, because of the increase of the probability to transit in the error state in Eq.~\eqref{eq:bit_error_against_tau}. Next, we can also see the retention accuracy decreases as the memory activation period $Q$ increases. This is because the probability of error after $M_{Q}$ refreshments $\bar{P}_{e}(M_{\zeta}, \tau)$ in Eq.~\eqref{eq:bit_error_after_refreshment} becomes larger. 
Next, Fig.~\ref{Fig:retention_analysis} (right) shows retention accuracy $\gamma(\tau, Q)$ against the retention accuracy threshold $\Delta$, where we set $Q = 50$ and $\tau = 128$~ms. From this figure, we can see that retention accuracy increases as $\Delta$ becomes larger as it allows more number of erroneous state for an memory array after $M_{\zeta}$ refreshments. Next, we can see that retention accuracy $\gamma(\tau, Q)$ increases as $\beta$ becomes larger. This is because higher $\beta$ reduces $P_e(\tau)$ in Eq.~\eqref{eq:bit_error_against_tau}, leading to the reduction of $\bar{P}_{e}(M_{\zeta}\tau)$ and increases $\gamma(\tau, Q)$. 

%
%
\subsubsection{Impact of $\zeta$}
\begin{figure}[t]
 \centering
 \begin{tikzpicture}
\definecolor{crimson2143940}{RGB}{214,39,40}
\definecolor{darkgray176}{RGB}{176,176,176}
\definecolor{darkorange25512714}{RGB}{255,127,14}
\definecolor{forestgreen4416044}{RGB}{44,160,44}
\definecolor{lightgray204}{RGB}{204,204,204}
\definecolor{mediumpurple148103189}{RGB}{148,103,189}

\def\hsep{1.4cm}
\def\vsep{2cm}
\def\vside{3.5cm}
\def\hside{0.14\textwidth}

\begin{groupplot}[
group style={group name=system, group size=2 by 1,  horizontal sep=\hsep, vertical sep=\vsep, xlabels at=edge bottom}, 
title style={at={(0.5,0.85)}},
anchor=south east,
height=\vside,
width=\hside,
scale only axis,
legend style={  
  at={(1.2, 1.0)}, 
  draw=none,
  fill opacity=0,
  anchor=south,  
  /tikz/every even column/.append style={column sep=0.3cm}
},
legend columns=2,
xmin=0, xmax=25,
xtick={0, 5, 10, 15, 20, 25},
xticklabels={
  \(\displaystyle {0}\),
  \(\displaystyle {5}\),
  \(\displaystyle {10}\),
  \(\displaystyle {15}\), 
   \(\displaystyle {20}\),
  \(\displaystyle {25}\), 
},
xlabel={$\zeta$},
xtick style={color=black},
scaled x ticks = false,
ymin=0.4, ymax=1,
ytick={0,0.25,0.5,0.75,1,1.5,2.},
yticklabels={
  \(\displaystyle {0.0}\),
  \(\displaystyle {0.25}\),
  \(\displaystyle {0.50}\),
  \(\displaystyle {0.75}\),
  \(\displaystyle {1.0}\),
  \(\displaystyle {1.5}\),
  \(\displaystyle {2.0}\),  
},
ylabel shift=-4pt,
xlabel shift=-3pt,
]

\nextgroupplot[
ylabel={Accuracy},
ymin=0.7,
ymax=1,
legend style={at={(0,0.4)},anchor=north west,legend cell align=left, align=left, draw=white!15!black,fill=white!15!black },
 legend image post style={xscale=0.8},
legend columns=1
]


\addplot[dashed, color=red, line width=1pt, mark size=1pt]
table {%
0.0	0.9478775733150987
5.0	0.936102848992003
10.0	0.9244743926435783
15.0	0.912990387299862
20.0	0.9016490385616547
25.0	0.8904485743201418

};
\addlegendentry{$\lambda = 0.0025$}
\addplot[    only marks, color=red, line width=1pt, mark size=1pt,mark=o, forget plot]
table {%
0.0	0.9504
5.0	0.9328
10.0	0.9271
15.0	0.9166
20.0	0.8994
25.0	0.888

};

\addplot[dotted,  color=blue, line width=1pt ]
table {%
0.0	0.9934389538520758
5.0	0.9689108586869869
10.0	0.9449883643493014
15.0	0.9216565185013151
20.0	0.898900737979857
25.0	0.8767067996813378

};
\addlegendentry{$\lambda = 0.0050$}

\addplot[    only marks, color=blue, line width=1pt, mark size=1pt,mark=x, forget plot]
table {%
0.0	0.992
5.0	0.9693
10.0	0.9446
15.0	0.9206
20.0	0.8964
25.0	0.8773

};

\addplot[ dashdotted, color=brown, line width=1pt, mark size=1pt]
table {%
0.0	0.9986968439051385
5.0	0.9619392250448323
10.0	0.9265344917498757
15.0	0.8924328502792791
20.0	0.8595863395796836
25.0	0.8279487638322259

};
\addlegendentry{$\lambda = 0.0075$}
\addplot[ only marks, color=brown, line width=1pt, mark size=1pt,mark=diamond, forget plot]
table {%
0.0	0.9989
5.0	0.9625
10.0	0.9269
15.0	0.8869
20.0	0.8636
25.0	0.836

};


\nextgroupplot[
ylabel={Energy $\bar{E}$},
ymin=0.35,
ymax=0.5,
  ytick={0.3, 0.4, 0.5},
 yticklabels={
 \(\displaystyle 0.3\),
  \(\displaystyle 0.4\),
  \(\displaystyle 0.5\),
  },
  legend style={at={(0,0.4)},anchor=north west,legend cell align=left, align=left, draw=white!15!black,fill=white!15!black },
 legend image post style={xscale=0.8},
legend columns=1
]


\addplot[dashed, color=red, line width=1pt, mark size=1pt]
table {%
0.0	0.4999594854162009
5.0	0.49377316429007206
10.0	0.4874669982994888
15.0	0.48138701830078034
20.0	0.47603362502648344
25.0	0.46961356411268823

};
\addlegendentry{$\lambda = 0.0025$}
\addplot[    only marks, color=red, line width=1pt, mark size=1pt,mark=o, forget plot]
table {%
0.0	0.4999594854162009
5.0	0.49377316429007206
10.0	0.4874669982994888
15.0	0.48138701830078034
20.0	0.47603362502648344
25.0	0.46961356411268823

};

\addplot[dotted,  color=blue, line width=1pt ]
table {%
0.0	0.49992026182683724
5.0	0.48744034472284914
10.0	0.4754151876497146
15.0	0.462906893124265
20.0	0.4531283682583556
25.0	0.4406101918211628

};
\addlegendentry{$\lambda = 0.0050$}
\addplot[ only marks, color=blue, line width=1pt, mark size=1pt,mark=x, forget plot]
table {%
0.0	0.49992026182683724
5.0	0.48744034472284914
10.0	0.4754151876497146
15.0	0.462906893124265
20.0	0.4531283682583556
25.0	0.4406101918211628

};
\addplot[ dashdotted, color=brown, line width=1pt, mark size=1pt]
table {%
0.0	0.49987971179358126
5.0	0.48178885608619815
10.0	0.4635735860659444
15.0	0.4458376420785612
20.0	0.4297055543433403
25.0	0.4145953247763683

};
\addlegendentry{$\lambda = 0.0075$}

\addplot[ only marks, color=brown, line width=1pt, mark size=1pt,mark=diamond, forget plot]
table {%

0.0	0.49987971179358126
5.0	0.48178885608619815
10.0	0.4635735860659444
15.0	0.4458376420785612
20.0	0.4297055543433403
25.0	0.4145953247763683

};




\end{groupplot}




\end{tikzpicture}
  \caption{Retrieval accuracy and normalized energy consumption obtained by analysis (line) and simulations (marker) against $\zeta$.} 
 \vspace{-4mm}
 \label{Fig:Impact_of_zeta}
\end{figure}
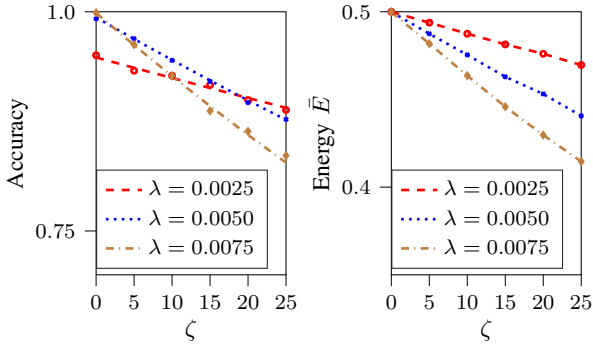

Fig.~\ref{Fig:Impact_of_zeta} shows the retrieval accuracy and normalized energy consumption, $\bar{E}$, against the time passed for sending the memory activation signals $\zeta$ for the different query arrival rate $\lambda$, where we set $\beta = 6400$, $\tau= 128$~ms, $N = 64$, $\Delta = 0.1$, $T_{\Sigma} = 0$, $w_{\mathrm{c}} = 0$, and $w_{\mathrm{l}} = 0$. From this figure, we can first see that the results of $\bar{E}$ and retrieval accuracy obtained by our theoretical analysis coincide with those of the simulation results, validating our analysis. 
Next, from Fig.~\ref{Fig:Impact_of_zeta}, we can see the trade-off between the retrieval accuracy and $\bar{E}$ through the value of $\zeta$. Specifically, the higher $\zeta$ leads to small $\bar{E}$ because of the shorter memory activation time, but the \gls{iot} device might fail to serve the query, deteriorating the retrieval accuracy. 
Next, from Fig.~\ref{Fig:Impact_of_zeta}, for the range of smaller $\zeta$ (e.g., $\zeta = 0$),  we can see that the retrieval accuracy becomes smaller as $\lambda$ becomes smaller. When the \gls{es} activates the memory earlier time, the model in the memory gradually degrade as per refreshment until the end of the episode. As the memory activation time ($t_q - \zeta -T_{\Sigma}$) becomes larger as $\lambda$ becomes smaller, the retrieval accuracy for $\lambda = 0.0025$ becomes smaller than that for $\lambda = 0.0075$. On the other hand, for the range of larger $\zeta$, we can see that the retrieval accuracy for the $\lambda = 0.0075$ becomes smaller than that for $\lambda = 0.0025$. This is because when the \gls{es} decides to activate memory later time, there is a risk of query arrival at the \gls{es} before memory activation, especially when the value of $\lambda$ is higher. In this case, the \gls{es} cannot serve the current tasks under the latency constraint, degrading the retrieve accuracy. 
On the other hand, from Fig.~\ref{Fig:Impact_of_zeta}, we can see that $\bar{E}$ becomes small as $\lambda$ increases due to the decreasing  memory activation time. This result shows the importance of memory activation timing $\zeta$ considering the trade-off between total energy consumption and retrieval accuracy.

\subsubsection{Impact of $\tau$}
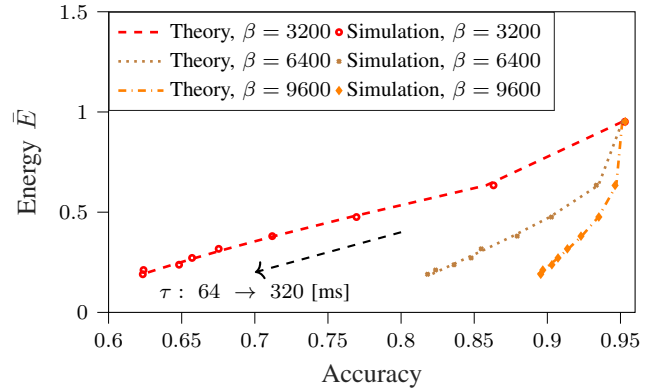
\begin{figure}[t]
 \centering
\begin{tikzpicture}
\definecolor{crimson2143940}{RGB}{214,39,40}
\definecolor{darkgray176}{RGB}{176,176,176}
\definecolor{darkorange25512714}{RGB}{255,127,14}
\definecolor{forestgreen4416044}{RGB}{44,160,44}
\definecolor{lightgray204}{RGB}{204,204,204}
\definecolor{mediumpurple148103189}{RGB}{148,103,189}

\def\hsep{1.2cm}
\def\vsep{1cm}
\def\vside{2.8cm}
\def\hside{0.18\textwidth}

\begin{axis}[%
width=7cm,
height=4cm,
scale only axis,
xmin=0.6,
xmax=0.96,
xlabel style={font=\color{white!15!black}},
xlabel={Accuracy},
ymin=0,
ymax=1.5,
ylabel style={font=\color{white!15!black}},
ylabel={Energy $\bar{E}$},
axis background/.style={fill=white},
legend style={at={(0,1)},anchor=north west,legend cell align=left, align=left, draw=white!15!black},
legend columns=2
]

\addplot[dashed, color=red, line width=1pt, mark size=1pt]
table {%
0.9512296226421596	0.9509821844061742
0.8581619521631337	0.6338602975871879
0.7652553537804373	0.47574588066461904
0.7131629674555102	0.3804179848033906
0.6809989942231053	0.31710108462265624
0.6593277204041967	0.2717968365499893
0.6437719360806378	0.23786968024321758
0.6320891117114096	0.21145443162437666
0.6229936862293751	0.19023728274665014

};
\addlegendentry{Theory, $\beta = 3200$}
\node[black, font=\footnotesize] at (axis cs:0.7,0.1) {$\tau:~64~\rightarrow~320$ [ms]};
\draw[->, thick, dashed, black] (axis cs:0.8,0.4) -- ++(0.5,-0.2);
\addplot[    only marks, color=red, line width=1pt, mark size=1pt,mark=o]
table {%
0.953	0.9509821844061742
0.8632	0.6338602975871879
0.7695	0.47574588066461904
0.7119	0.3804179848033906
0.6753	0.31710108462265624
0.6571	0.2717968365499893
0.6482	0.23786968024321758
0.6239	0.21145443162437666
0.6234	0.19023728274665014

};
\addlegendentry{Simulation, $\beta = 3200$}

\addplot[dotted,  color=brown, line width=1pt ]
table {%
0.9512296226421596	0.9512740633560088
0.9349529395081246	0.6344575115527387
0.9016490385616547	0.47562037223317283
0.8762932809005646	0.3806166303569464
0.8581859714742736	0.31709003335609887
0.8449277130037433	0.27163129828551497
0.8348796722495442	0.23769286651261137
0.8270801077481053	0.21138657155110566
0.8208185723725665	0.1901930665256872

};
\addlegendentry{Theory, $\beta = 6400$}

\addplot[    only marks, color=brown, line width=1pt, mark size=1pt,mark=x]
table {%
0.9521	0.9512740633560088
0.9335	0.6344575115527387
0.9028	0.47562037223317283
0.8794	0.3806166303569464
0.8548	0.31709003335609887
0.8479	0.27163129828551497
0.8362	0.23769286651261137
0.8236	0.21138657155110566
0.8181	0.1901930665256872

};
\addlegendentry{Simulation, $\beta = 6400$}

\addplot[ dashdotted, color=orange, line width=1pt, mark size=1pt]
table {%
0.9512296226421596	0.951684016363628
0.947127598566313	0.6334912006287624
0.9349250284711721	0.4759807161204301
0.9234708171521493	0.38036736344656935
0.9143247223807629	0.317118875755848
0.9072639863260599	0.27160921978405705
0.9016765437254549	0.23760906481507163
0.8971682760975684	0.21125081486222402
0.8935377286909627	0.19015811336580238

};
\addlegendentry{Theory, $\beta = 9600$}

\addplot[ only marks, color=orange, line width=1pt, mark size=1pt,mark=diamond]
table {%
0.9533	0.951684016363628
0.9463	0.6334912006287624
0.9352	0.4759807161204301
0.923	0.38036736344656935
0.9137	0.317118875755848
0.9073	0.27160921978405705
0.9031	0.23760906481507163
0.8969	0.21125081486222402
0.8955	0.19015811336580238

};
\addlegendentry{Simulation, $\beta = 9600$}


\end{axis}

\end{tikzpicture}
  \caption{The set of retrieval accuracy and normalized energy consumption for the different value of $\beta$ where we vary the value of $\tau$.} 
 \vspace{-3mm}
 \label{Fig:Region_tau}
\end{figure}

Fig.~\ref{Fig:Region_tau} shows the set of normalized energy consumption and retrieval accuracy when we change the value of $\tau$ from [$64$, $320$]~ms with the interval of $32$~ms, where we apply $\lambda= 0.0025$, $\zeta = 20$, $N = 64$, $\Delta = 0.1$, $T_{\Sigma} = 0$, $w_{\mathrm{c}} = 0$, and $w_{\mathrm{l}} = 0$. From Fig.~\ref{Fig:Region_tau}, we can see the trade-off between retrieval accuracy and normalized energy consumption, through the value of $\tau$. For the smaller (larger) range of accuracy, each node applies relatively large (small) refresh period $\tau$, which decreases (increases) the energy consumed by the refreshment, but the quality of retained data at the memory becomes small. 
Next, from Fig.~\ref{Fig:Region_tau}, we can see that, as $\beta$ becomes smaller, more energy is needed to reach the same level of retrieval accuracy. As the small $\beta$ corresponds to the fast degradation of stored data in the memory. This requires for \gls{iot} device to apply smaller $\tau$ to realize the target accuracy level, which increases the total energy consumption due to the increasing number of refreshments. 
\subsection{Energy efficiency}\label{sec:energy_efficiency}
To investigate the system level performance of our proposed scheme, we consider the six parameters settings for ($T_{\Sigma}, w_{\mathrm{c}}, w_{\mathrm{l}}, \beta$) listed in Table~\ref{tab:parameters}. These represent a wide range of values that allow us to compare our proposed scheme with the baseline in terms of retrieval accuracy and normalized energy consumption under diverse conditions.

\begin{table}
    \centering
    \renewcommand{\arraystretch}{1.2}
    \caption{Parameter settings for the evaluation of energy efficiency.}
    
    \begin{tabular}{lcccccc}
    \toprule
    Parameter & \multicolumn{6}{c}{Setting}\\\cmidrule{2-7}
    & $1$ & $2$ & $3$ &$4$ & $5$ & $6$\\\midrule
       $T_{\Sigma}$  &  $0$ & $0$ & $0$ & $100$ & $0$ & $0$\\
         $w_{\mathrm{c}}$ & $0$ & $0$ & $0$ & $0$ & $500$ & $10^5$\\
         $w_{\mathrm{l}}$ &$0$& $0$ & $500$ & $500$ & $500$ &$500$\\ 
         $\beta$ & $3200$ & $6400$ & $3200$ & $6400$ & $3200$ &$3200$\\\bottomrule
    \end{tabular}
    \label{tab:parameters}
\end{table}

\begin{figure}[t]
 \centering
  \begin{tikzpicture}
\definecolor{crimson2143940}{RGB}{214,39,40}
\definecolor{darkgray176}{RGB}{176,176,176}
\definecolor{darkorange25512714}{RGB}{255,127,14}
\definecolor{forestgreen4416044}{RGB}{44,160,44}
\definecolor{lightgray204}{RGB}{204,204,204}
\definecolor{mediumpurple148103189}{RGB}{148,103,189}

\def\hsep{1.2cm}
\def\vsep{1cm}
\def\vside{2.8cm}
\def\hside{0.18\textwidth}

\begin{axis}[%
width=7cm,
height=4cm,
scale only axis,
xmin=0.6,
xmax=0.95,
xlabel style={font=\color{white!15!black}},
xlabel={$U_{\mathrm{th}}$},
ymin=0,
ymax=1.4,
ytick={0,0.25, 0.5,0.75, 1,1.25},
yticklabels={
  \(\displaystyle {0}\),
  \(\displaystyle {0.25}\),
  \(\displaystyle {0.5}\),
  \(\displaystyle {0.75}\),
  \(\displaystyle {1}\),
  \(\displaystyle {1.25}\),
},
scaled y ticks = false,
ylabel style={font=\color{white!15!black}},
ylabel={Minimum $\bar{E}$},
axis background/.style={fill=white},
legend style={at={(0,1)},anchor=north west,legend cell align=left, align=left, draw=white!15!black},
legend image post style={xscale=0.8},
legend columns=3
]

\addplot[solid, color=red, line width=1pt, mark= square,mark size=1pt, mark options={solid, red}]
table {%
0.6	0.10775828492818908
0.65	0.1984420942506346
0.7	0.3262725195779148
0.75	0.46615961420027746
0.8	0.49746643909727967
0.85	0.8500160902253981
0.9	0.9003245225862656
0.95	0.9500411305585277

};
\addlegendentry{Setting 1}


\addplot[dotted,  color=brown, line width=1pt , mark=o,mark options={solid, brown}]
table {%
0.6	0.07962997691076046
0.65	0.08626212418668236
0.7	0.09298057648246677
0.75	0.09959785381465606
0.8	0.10628674850760782
0.85	0.1661840272855016
0.9	0.3328636616872946
0.95	0.9500411305585277

};
\addlegendentry{Setting 2}

\addplot[ dashed, color=blue, line width=1pt, mark size=1pt,mark=diamond,mark options={solid, blue}]
table {%
0.6	0.18540013879475564
0.65	0.2778506911545259
0.7	0.40459672058102486
0.75	0.5407570874314825
0.8	0.5770738058522185
0.85	0.9180228178365689
0.9	0.9723562467774461
0.95	1.0260505015907517

};
\addlegendentry{Setting 3}

\addplot[ dashdotted, color=orange, line width=1pt, mark size=1pt,mark=triangle  ,mark options={solid, orange}]
table {%
0.6	0.13700486028815903
0.65	0.1484155933587772
0.7	0.3210641088922804
0.75	0.8101522932427543

};
\addlegendentry{Setting 4}

\addplot[ dashed, color=cyan, line width=1pt, mark size=1pt,mark=pentagon,mark options={solid, cyan}]
table {%
0.6	0.24355595724079226
0.65	0.33079367672774657
0.7	0.4471463515693029
0.75	0.5697693647632007
0.8	0.6080345194246984
0.85	0.9129849092895034
0.9	0.9670201682494889
0.95	1.020419760729793

};
\addlegendentry{Setting 5}

\addplot[ dotted, color=black, line width=1pt, mark size=1pt,mark=oplus,mark options={solid, black}]
table {%
0.6	0.6033214536574977
0.65	0.6535703283980241
0.7	0.7035930572359673
0.75	0.7536666378145013
0.8	0.804282153921038
0.85	0.8540161847675782
0.9	0.9045613638065693
0.95	0.9545119333879026

};
\addlegendentry{Setting 6}

\addplot[dash dot,  color=black, line width=1pt ]
table {%
0.6 1
0.95 1
};

\end{axis}

\end{tikzpicture}
  \vspace{-4mm}
  \caption{Maximum Energy consumption against $U_{\mathrm{th}}$.}
 \label{Fig:minimum_energy}
 \vspace{-4mm}
\end{figure}
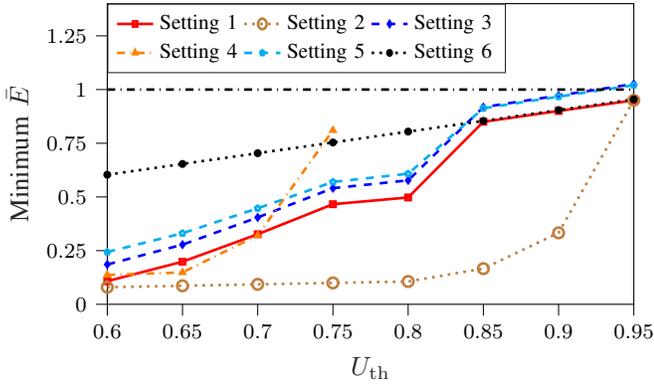
Fig.~\ref{Fig:minimum_energy} shows the minimum $\bar{E}$ against the retrieval accuracy constraint $U_{\mathrm{th}}$ with different settings, where we set $\lambda = 0.0025$, $N = 64$, and $\Delta = 0.1$. First, from Fig.~\ref{Fig:minimum_energy}, it can be observed that setting 3 always requires higher energy than setting 1. This is because of the additional energy required for model loading from the storage to the memory. 
Next, compared with settings 1 and 2, we can see that the minimum energy consumption of setting 2 is significantly smaller than that of setting 1, especially when $U_{\mathrm{th}}$ is small. In setting 2, as the accuracy degradation due to the lower refresh period becomes smaller for the larger $\beta$, the \gls{iot} device can assign the large $\tau$ to reach the retrieval accuracy requirement $U_{\mathrm{th}}$, reducing energy consumed by refreshments. Next, from Fig.~\ref{Fig:minimum_energy}, we can also see that the normalized energy consumption of setting 4 is smaller than that of settings 1 and 3 for the smaller range of $U_{\mathrm{th}}$ because of the larger $\beta$ mentioned above. We can also see that the normalized energy consumption of setting 4 becomes highest when $U_{\mathrm{th}} = 0.75$. This is because, in order to maintain the high retrieval accuracy, the \gls{es} requires to assign relatively small $\tau$ for memory refreshment, increasing the energy consumption at the memory in addition to the additional energy cost $E_{\mathrm{l}}$. 
Here, in setting 4, there are no values plotted for the range of $U_{\mathrm{th}} > 0.75$. This is because our proposed scheme can not provide retrieval accuracy higher than $U_{\mathrm{th}}$ due to the large overhead, $T_{\Sigma}$. 
Next, compared with the results of settings 3 and 5, we can see that $\bar{E}$ increases as $w_c$ increases. However, we can see that, even for the extreme case (setting 6), our proposed scheme can still offer smaller energy consumption than baseline scheme, i.e., $\bar{E} < 1$, thanks to the reduction of energy consumption required by \gls{dram} refreshment by activating memory at the right time. 
Finally, we can see that in all settings, normalized energy consumption becomes larger as $U_{\mathrm{th}}$ becomes larger, while offering smaller energy consumption than always-on approach. The maximum retrieval accuracy constraint $U_{\mathrm{th}}$ under which our proposed scheme can achieve lower energy consumption than always-on approach are 0.95, 0.95, 0.9, 0.75, 0.9, and 0.95 for setting 1, 2, 3, 4, 5, and 6, respectively, indicating the efficiency of our proposed scheme. 
\section{Conclusion}
We have proposed wireless memory activation and memory approximation for pull-based communication system, in which the \gls{es} remotely activates/approximates the \gls{dram} memory for relevant \gls{ml} model at the right timing. We analyzed the total energy consumption of the \gls{iot} device and retrieval accuracy of the proposed scheme considering the energy consumed by memory refreshment. 
The numerical results showed that our proposed scheme can realize high energy efficiency under the retrieval accuracy constraint, compared with the always-on approach. 
Future work includes the design of remote memory activation protocol for more general case considering the multiple tasks with different requirements. 
\section*{Appendix-Proof of Eq.~\eqref{eq:PMF_bit_error}}
The Eq.~\eqref{eq:PMF_bit_error} can be calculated as follows:
\begin{equation*}\begin{split}
& \Pr(N_e(M_{\zeta}) = k) 
\\&= \sum_{ m = 0}^{\infty}\int_{m\tau +\zeta+T_{\Sigma}}^{(m+1)\tau +\zeta+T_{\Sigma}}\binom{N}{k}[1-q_e^{m}]^k q_e^{m(N -k)}\lambda\,e^{-\lambda\,t}dt
\\&\overset{(a)}{=}e^{-\lambda\,\left(\zeta +T_{\Sigma}\right)}\sum_{m=0}^{\infty}\psi^m(1-\psi)\binom{N}{k}[1-q_e^{m}]^k q_e^{m(N -k)}
%
%
\\&\overset{(b)}{=}e^{-\lambda\,\left(\zeta +T_{\Sigma}\right)}(1-\psi)\binom{N}{k}\sum_{j=0}^{k}\binom{k}{j}(-1)^{j}\sum_{m= 0}^{\infty}\psi^mq_e^{mc}
\\&\overset{(c)}{=}e^{-\lambda\,\left(\zeta +T_{\Sigma}\right)}(1-\psi)\binom{N}{k}\sum_{j=0}^{k}\binom{k}{j}(-1)^{j}\frac{1 }{1-\psi\,q_e^{c}}
\end{split}\end{equation*}
where (a) $q_e = 1- P_e$ and $\int_{m\tau +\zeta+T_{\Sigma}}^{(m+1)\tau +\zeta+T_{\Sigma}}\lambda\,e^{-\lambda\,t}dt=e^{-\lambda\,\left(\zeta +T_{\Sigma}\right)}\psi^m(1-\psi)$, with $\psi = \exp(-\lambda\,\tau)$; (b), we apply the binomial theorem $[1-q_e^{m}]^k =\sum_{j=0}^k\binom{k}{j}(-1)^{j}q_e^{mj}$; and (c) we put $c= N-k +j$ and use $\sum_{m= 0}^{\infty}\psi^{m}q_e^{mc} = \frac{1 }{1-\psi\,q_e^{c}}$ where $0\leq \psi\,q_e^{N-k +j}\leq 1$.

\section*{Appendix-Proof of Eq.~\eqref{eq:proposed_ene}}
The Eq.~\eqref{eq:proposed_ene} can be calculated as follows: 
\begin{equation*}\begin{split}
 &\mathbb{E}[E_\mathrm{tot}^\mathrm{p}]\\&=\left(\int_{t=\zeta +T_{\Sigma}}^{\infty}\left(\lfloor\frac{t-\zeta-\,T_{\Sigma}}{\tau}\rfloor +w_{c}+w_{l}\right)\lambda{e^{-\lambda\,t}}dt\right) I_{\mathrm{M}}\xi_{\mathrm{re}}
    \\&=\Bigg(\sum_{m=0}^{\infty}\int_{m\,\tau + \zeta + T_{\Sigma}}^{(m+1)\,\tau + \zeta + T_{\Sigma}}m  \lambda{e^{-\lambda\,t}}dt\\&+\int_{t=\zeta +T_{\Sigma}}^{\infty}(w_{c} + w_{l})\lambda{e^{-\lambda\,t}}dt\Bigg) I_{\mathrm{M}}\xi_{\mathrm{re}} 
    \\&=e^{-\lambda\,\left(\zeta +T_{\Sigma}\right)}\left((1-\psi)\sum_{m=0}^{\infty}m\psi^m +(w_{c} +w_{l})\right) I_{\mathrm{M}}\xi_{\mathrm{re}}
\end{split}\end{equation*}
\bibliographystyle{IEEEtran}
\bibliography{IEEEabrv,Ref_2}

\end{document}